\documentclass[fleqn,twoside]{article}
\usepackage{espcrc2}

\usepackage{graphicx}
\usepackage[figuresright]{rotating}
\input{epsf}
\newcommand{\beq}{\begin{equation}}
\newcommand{\eeq}{\end{equation}}
\newcommand{\bea}{\begin{eqnarray}}
\newcommand{\eea}{\end{eqnarray}}

\newcommand{\tx}{\textstyle}

\newcommand{\tr}{{\rm tr}}

\newcommand{\vev}[1]{\Big\langle #1 \Big\rangle}

\newcommand{\V}{{\cal V}}
\newcommand{\cO}{{\cal O}}

\newcommand{\AmS}{{\protect\the\textfont2
  A\kern-.1667em\lower.5ex\hbox{M}\kern-.125emS}}

\title{Vortex waistlines and vortex gas}

\author{Tam\'as G. Kov\'acs\address{NIC/DESY Zeuthen, Platanenallee 6,
D-15738, Zeuthen, Germany}%
        \thanks{Supported by the European 
Community's Human Potential Programme under contract 
HPRN-CT-2000-00145, ``Hadrons/Lattice QCD'' and 
by Hungarian science grant OTKA-T032501. On leave
from Department of Theoretical Physics, University of P\'ecs,
Hungary.},
        E.T. Tomboulis\address{Department of Physics, UCLA, Los Angeles, 
CA 90095-1547, USA}\thanks{Research supported by 
NSF grant NSF-PHY9819686.}}
       
\begin{document}

\begin{abstract}
We investigate the manner in which a linear potential arises from 
fluctuations due to vortices linked with the Wilson loop. 
In particular, the customary naive picture is critically  
reexamined.  
\vspace{1pc}
\end{abstract}

\maketitle

A large body of work over the last few years \cite{procs} supports the 
picture of long thick vortices as the most important long range 
degrees of freedom for confinement.  

In this note we wish to examine more closely the manner in which a linear 
potential arises from vortex fluctuations. 
One customarily assumes a picture of randomly distributed 
vortices of certain thickness and 
basically arbitrary length. Furthermore, in the presence of a Wilson loop, 
one assumes that the free energy cost and hence the probability 
for a vortex to pierce anywhere inside the area enclosed by 
the loop is constant for all loop sizes. Summing then 
{\it independently} over all   
vortices linking with the loop one gets 
area law behavior for the loop expectation: 
\bea
W[C] &\sim& 1 + (-1)\, k|A_C| + {((-1)\,k|A_C|)^2\over 2!} + \cdots 
\nonumber\\
 &=& \exp(\,- k|A_C|\,)\,. \label{naive}
\eea 
How accurate is this picture? 

We consider explicitly only the case of $SU(2)$ for 
simplicity. 
We work on a lattice $\Lambda$ of linear extension 
$L_\mu$ in spacetime direction $\mu=1, \ldots, d $. 
Let $\cO[\V_{\mu\nu}] $ denote the operator that flips
the sign of the coupling (introduces a $Z(2)$ `twist') 
on a coclosed set of
plaquettes $\V_{\mu\nu}$ winding around the periodic lattice
perpendicularly to the $\mu\nu-$directions. The twist introduces 
a discontinuous gauge transformation with multivaluedness in $Z(2)$, 
i.e. a vortex trapped in the periodic lattice. 
The expectation of this operator 
defines the vortex free energy 
\beq
\exp (-F_v[\mu\nu]) = \vev{\cO[\V_{\mu\nu}]}\;. \label{vfe} 
\eeq
The expectation 
depends only on the directions in which $\V$ winds through the lattice, 
not its exact shape or location, nor  
the number mod 2 (mod $N$) 
of homologous sets $\V$ carrying a twist. This expresses the 
mod 2 conservation of flux, a simple consequence of which is  
\beq
\langle \cO[\V_{\mu\nu}] \, \cO[\V^{\ \prime}_{\mu\nu}] \rangle = 1.
   \label{mod2}
\eeq

Our argument in the following 
assumes that the vortex free energy 
(\ref{vfe}) behaves as follows.  
For sufficiently large $|A_{\mu\nu}|$ 
and dimension $d\leq 4$ 
\beq
F_v[\mu\nu] \sim  (\prod_{\lambda\not= \mu\nu} L_\lambda\,) 
\;\exp(\,- \rho(\beta)\,|A_{\mu\nu}|\,),\label{vfe1}
\eeq
where $|A_{\mu\nu}| = L_\mu L_\nu$. 

This is the optimal  
behavior  
under exponential transverse spreading 
of the flux introduced by the twist on $\V$, with 
$\rho$ approaching, at least asymptotically, the exact 
linear potential string tension. It is dictated by physical reasoning, 
and explicitly seen in strong coupling expansion. 
Recently, it has become possible to exhibit 
this behavior at {\it large} $\beta$ in extensive numerical 
simulations \cite{vfe}. 

The $Z(2)$ Fourier transform of (\ref{vfe}): 
\beq
\exp(-F_{\rm el}) = \vev{\,\frac{1}{2}(\, 1 - \cO[\V_{\mu\nu}]\,)} 
\label{efe}
\eeq 
gives the dual (w.r.t. the center) color-electric     
free energy. 
The mod 2 conservation of the magnetic flux 
is now expressed by the projection property
\bea
\vev{\,\frac{1}{2}(\, 1 - \cO[\V_{\mu\nu}]\,)
\frac{1}{2}(\, 1 - \cO[\V_{\mu\nu}^{\ \prime}]\,)}
   \qquad\quad & &    \nonumber \\
 = \vev{\,(\, 1 - \cO[\V_{\mu\nu}]\,)/2}\;. \hspace{3cm}& &   
\label{proj} 
\eea   

There are several identities involving the Wilson loop operator 
and the operator $\cO[\V]$ that can serve as the 
starting point for deriving rigorous inequalities relating the 
Wilson loop to the flux order parameters (\ref{vfe}), (\ref{efe}) 
\cite{KT1}. A simple such relation, which is easily 
obtained and the only one needed here, is : 
\beq
W[C] = \vev{\frac{1}{2}\tr U[C]\,\frac{1}{2}\Big(
\,1-\cO[\V^{\prime}]\,\Big)}\;. \label{rel}  
\eeq    
In (\ref{rel}), $W[C] = \vev{\frac{1}{2}\tr U[C]}$ is the Wilson loop 
expectation for a loop $C$, and $\V^{\prime}$ is a coclosed 
set of plaquettes linking with $C$.

\begin{figure}[hbt]
\vspace{-0.3cm}
{\hfill\epsfysize=4.5cm\epsfbox{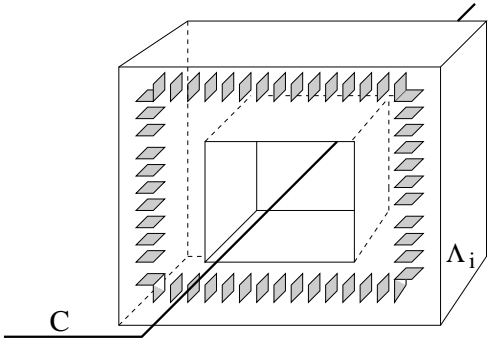}\hfill}
\vspace{-0.3cm}
\caption[vg7]{\label{vg7}Vortex container $\Lambda_i$  
enclosing coclosed plaquette set $\V_i$ (shaded) linking with 
loop $C$.}
\vspace{-0.7cm}
\end{figure}

Now by the property (\ref{proj}), one may insert multiple 
$\frac{1}{2}(\,1-\cO[\V]\,)$ factors in (\ref{rel}) 
corresponding to a collection of coclosed sets $\{\V_i\}$ linking with 
the loop $C$.
Imagine enclosing each such $\V_i$ in a `vortex container' \cite{MP}, 
i.e. a sublattice $\Lambda_i\subset \Lambda$ containing $\V_i$ and 
wrapping around $C$ (figure \ref{vg7}). 

Furthermore, imagine integrating over 
the bond variables 
in the interior of each container, keeping bond variables on its 
boundary $\partial\Lambda_i$ fixed, and define  
\beq 
f_{\Lambda_i}(U_{\partial\Lambda_i}) = \frac{1}{2}\,\Bigg(\, 1 - 
{z^{(-)}_{\Lambda_i}(U_{\partial\Lambda_i})\over 
z^{(+)}_{\Lambda_i}(U_{\partial\Lambda_i})} \,\Bigg) .\label{vcon}
\eeq
In (\ref{vcon}),  
$z^{(\mp)}_{\Lambda_i}(U_{\partial\Lambda_i})$ denotes  
the partition function for (the interior of) $\Lambda_i$ 
with $(-)$ or without $(+)$ twist on $\V_i$.  
Note that $f_{\Lambda_i}(U_{\partial\Lambda_i})$  
is nothing but the electric free energy   
now defined on a lattice $\Lambda_i$ (the vortex container) 
with fixed (instead of periodic) b.c.  
in the directions transverse to $\V_i$.
By these steps one then obtains from (\ref{rel}) the exact relation   
\beq 
W[C] = \vev{\;{\tx\frac{1}{2}}\tr U[C]\;
\prod_i\,f_{\Lambda_i}(U_{\partial\Lambda_i})\;}, \label{W5}
\eeq
which, in turn, immediately yields the bound:  
\beq
W[C] \leq \prod_i\; \max_{U_{\partial\Lambda_i}}
|f_{\Lambda_i}(U_{\partial\Lambda_i})|. \label{ineq}
\eeq 
In (\ref{ineq}) the maximum is taken over all values of the bond variables 
on the boundary $\partial\Lambda_i$.

The bound given by (\ref{ineq}) leads to area law according to the 
naive notion of independent vortices linking with the loop {\it provided}  
they grow long enough to be able to pierce at any point of the area spanning 
the loop. For this to happen one must assume, according to (\ref{vfe1}), 
that they grow thick enough to reach the regime:  
\beq 
 -\ln(z^{(-)}_{\Lambda_i}(U_{\partial\Lambda_i})/ 
 z^{(+)}_{\Lambda_i}(U_{\partial\Lambda_i})) 
 \sim |\V_i|\exp(-\rho\,d_i^2)
     \label{eq:free_E}
\eeq
for all $U_{\partial\Lambda_i}$. Here $d_i$ denotes the size of $\Lambda_i$ 
in each of the two directions transverse to the enclosed 
set $\V_i$; its longitudinal size is given by $|\V_i|$. 
It follows that to keep the free energy cost
of each vortex less than a fixed value $f$, we need: 
\beq
d^2_i \geq \frac{1}{\rho} \; \ln\left( |\V_i|\,/\,f \right). 
\eeq
Then also $\max f_{\Lambda_i}(U_{\partial\Lambda_i}) < \frac{1}{2}
(1-e^{-f}) < {\rm const}$.   
But, $|\V_i|$ (in $d$ dimensions) is of order $R^{d-2}$ 
in order to have linkage through points away from the perimeter of   
a rectangular loop of side lengths $T$ and $R$, $T >  R$.    
Such a loop can then accommodate $\sim RT/\ln R$ 
containers wrapped around it. Thus we obtain  
a confining but not quite purely linear potential 
\beq 
V(R)  \geq \mbox{const}\,R/\ln R \,. \label{quasilin}
\eeq

By the same reasoning we now see that the familiar argument for area law  
summarized in (\ref{naive}) must also fail in the same manner. 
It ignores the actual free energy requirements for having vortices of 
sufficient length link with a large loop:  
the type of discussion just given above for (\ref{ineq}) applies 
to each term in the summation in (\ref{naive}). Thus, for 
one vortex to link {\it anywhere} with 
a large loop of side lengths $T$ and $R$ ($T \gg R$), at bounded fixed cost,  
a vortex cross section area of order $\ln R$ is required.   
This leads at best to (\ref{quasilin}), {\it not} (\ref{naive}).  

The discrepancy comes from the treatment of vortices 
as localized and independent. 
For each vortex enclosed in a vortex container of fixed, but 
sufficiently large, width $d$, the vortex bulk free 
energy cost inside is correctly estimated from (\ref{eq:free_E}).   
In addition, however, the vortex is surrounded by the pure 
gauge long tail that encodes its nontrivial topology and flux quantization. 
The tail incurs no additional action cost, but is of infinite range,     
and communicates the presence of a vortex inside 
the container to other vortices, or other topological 
obstructions, outside  
enforcing flux conservation mod N. 
This long distance interaction allows a system of vortex excitations  
to adjust the amount of flux spreading, i.e adjust the 
thickness of vortices to minimize the free energy of the 
system. The thickness of 
vortex cores then is not fixed, but is adjusted relative to 
their length as required by the presence of other vortices and/or  
other obstructions (e.g. Wilson loop legs) sensitive   
to the presence of topological $Z(N)$ flux.  

Thus, in the presence of the Wilson loop source, the optimal 
configurations for the system are not those of multiple isolated  
linked vortices, each of some fixed free energy 
cost, hence length 
$|\V| \sim R^{(d-2)}$ and fixed width $d^2 \sim \ln R \ll R,T$. 
It is more advantageous, in terms of free energy cost, for 
multiple linking vortices to thicken and merge,  
the total topological flux being conserved mod N. 
Since the Wilson loop operator is affected   
by the topological flux through it only mod N, this should 
optimize the expectation. 

The exact expression 
(\ref{W5}) holds for any number of factors in the product 
inside the expectation. In view of the above discussion, 
optimization requires combining containers into ones as thick as 
possible by integrating over the boundary fields of neighboring 
containers. With $T\gg R$, this amounts to taking 
containers in the product in (\ref{W5}), 
(\ref{ineq}) having   
transverse area $\sim R^2$, and longitudinal extension  
$\sim (\mbox{const}\,R)^{(d-2)}$. 
(\ref{ineq}) now gives
\beq
V(R)  \geq \mbox{const}\,R - \mbox{const}\,(\,\ln R + 
\mbox{const}\,)/R \label{strictlin}
\eeq 
replacing (\ref{quasilin}). For loops with $T
\stackrel{\ }{\stackrel{\tx >}{ \sim}}R$, 
strict linear potential arises  
essentially from thick vortex fluctuations spanning the 
entire loop area.

Our discussion shows that an effective picture of the 
long distance confining fluctuations as isolated, 
independent vortices winding over long distances -  
i.e., as some kind of an approximately dilute or weakly 
interacting vortex gas - is not generally applicable. 
In general 
vortices cannot be considered isolated, and a definite 
number of vortices, specified more precisely than mod N, 
cannot necessarily be unambiguously assigned to every configuration. 
This accords with the picture of `condensation'  of vortex flux 
and `percolated' vortices over distances 
$\geq$ 1 fm, as suggested 
by the simulations. The effective description of this vacuum is 
discussed in \cite{KT1}


\begin{thebibliography}{9}
\bibitem{procs} R.W.\ Haymaker, Phys.\ Rept.\ {\bf 315} (1999) 153; 
Nucl. Phys. {\bf B} (Proc. Suppl.) {\bf 83-84} 
(2000) Lattice 99; Nucl. Phys. {\bf B} (Proc. Suppl.) 
{\bf94} (2001) Lattice 2000.    
\bibitem{vfe} T.G. Kov\'acs and E.T. Tomboulis, Phys. Rev. Lett.  
85 (2000) 704; 
Ch. Hoebling, C. Rebbi 
and V.A. Rubakov, Phys. Rev. {\bf D63} (2001) 034506;   
Ph. de Forcrand, M. D'Elia and M. Pepe, Phys. Rev. Lett. {\bf 86} 
(2001) 1438; Ph. de Forcrand and L. von Smekal, hep-lat/0107018.   
\bibitem{KT1} T.G. Kov\'acs and E.T. Tomboulis, hep-lat/0108017.  
\bibitem{MP} G. Mack and V.B. Petkova, Ann. Phys.
{\bf 125} (1980) 117.  
 

\end{thebibliography}
\end{document}